\documentclass[aps,
12pt,
final,
notitlepage,
oneside,
onecolumn,
nobibnotes,
nofootinbib,
superscriptaddress,
noshowpacs,
centertags]
{revtex4}
\begin{document}
\title{Phenomenology of Vector-Meson Electroproduction \\on Spinless Targets}
\author{\firstname{S.~I.}~\surname{Manaenkov}}
\email[]{Email:  manaenkov_si@pnpi.nrcki.ru}
\affiliation{B.~P.~Konstantinov Petersburg Nuclear Physics Institute of Research Center "Kurchatov Institute",\\
Gatchina, Russia}
\begin{abstract}
 Advantages of the amplitude method for the vector-meson-production 
description in respect of the spin-density-matrix-element (SDME) method
are discussed. 
It is shown that for any nonzero amplitudes the angular distribution of
final particles is non-negative. The exact formula for $R = \frac{d \sigma_L}{d t}/\frac{d \sigma_T}{d t}$ in terms of the SDMEs for spinless targets and the new approximate
formula for nucleons are presented.
\end{abstract}

\maketitle

\section{\label{sec:intro}Introduction}
In the recent time, the quark-gluon 
structure of nucleons and nuclei is studied in a vector-meson (VM) production in exclusive deep-inelastic scattering (DIS) of leptons and described in 
QCD-motivated models \cite{INS}.
Indeed, the VM production 
is one of two basic processes from which the generalized parton distributions can be extracted. 
As was shown in \cite{R83,R99}
 the leading amplitudes of the VM production on the nucleon by the photon with a large virtuality, 
$Q^2$ and small Bjorken $x_B$ depend on  the gluon distributions $G(x_B,Q^2)$ and $\Delta G(x_B,Q^2)$ in the nucleon for DIS of unpolarized and 
polarized particles, respectively.

The main phenomenologic method of the vector-meson-production description is the spin-density-matrix-element (SDME) method \cite{SW} when the SDMEs are 
 obtained from 
the angular distribution of the final particles. 
Another method is the amplitude method  in which the helicity-amplitude ratios (HARs) are considered as free parameters.
This method was independently proposed by the H1 \cite{AMA} 
and HERMES \cite{DC-84} collaborations.
One of advantages of the amplitude method is the following. Now, the QCD-motivated models applied for the VM-production description can
predict not all the helicity amplitudes. Any SDME contains, as a rule, contributions of many amplitudes into the numerator of the expression for 
it.
Therefore it is sometimes difficult to establish which amplitudes are predicted by the 
model wrong. 
In the amplitude method, the extracted HARs can be directly compared with the 
 calculated ones in the model. 
It will be shown that for spinless targets,
all the HARs can be extracted from the data.
The knowledge of the helicity amplitudes for the spinless nuclei helps to 
understand how the quark and gluon 
distributions in nuclei are related to those in nucleons. 
This problem is not trivial as it is 
learned from the observed EMC effect \cite{ABB}.
The problem, how the quark distributions in nucleons and nuclei are related, is not fully solved up to now
(see, for instance, review \cite{HMPW}).
 Another advantage of the amplitude method 
compared with the SDME method consists in reduction of the free-fit-parameter number.

\section{\label{sec:solution}Solution of equations for helicity amplitude ratios}

In the present paper, exclusive DIS is considered in the one-photon-exchange approximation. 
The SDMEs in the Schilling-Wolf (SW) representation \cite{SW}
of the subprocess
\begin{equation}
\gamma^* (\lambda_{\gamma}) + N(\lambda_{N}) \to V(\lambda_{V})+N (\mu_{N})
\label{gnvn}  
\end{equation}
were written  for lepton scattering from the nucleon.
Here, $\gamma^*$ denotes the virtual photon emitted by the lepton, $N$ is the nucleon, while $V$ denotes the produced vector meson.
The helicities of the particles in Eq.~$(\ref{gnvn})$ are given in parentheses. 
The formulas for the SDMEs for the spinless target are obtained after the following reduction. First, all the unnatural-parity-exchange (UPE) amplitudes are
to be put equal to zero. Second, the substitution should be performed
for the sum with the natural-parity-exchange (NPE) amplitudes 
\begin{equation}
\widetilde{\sum}\{T_{\lambda_V  \lambda_{\gamma} } T^*_{\lambda'_V \lambda'_{\gamma} }\}
\equiv \frac{1}{2}\sum_{\lambda_N,\mu_N}T_{\lambda_V \mu_N \lambda_{\gamma} \lambda_N} T^*_{\lambda'_V \mu_N \lambda'_{\gamma} \lambda_N}
\to F_{\lambda_V \lambda_{\gamma}} F^*_{\lambda'_V \lambda'_{\gamma} }.
\label{t-amp-f}
\end{equation}
Here, the NPE amplitude $T_{\lambda_V \mu_N \lambda_{\gamma} \lambda_N}$ describes the reaction in Eq.~$(\ref{gnvn})$, while $F_{\lambda_V 
\lambda_{\gamma}}$ is the helicity amplitude of the subprocess $\gamma^* (\lambda_{\gamma}) + S \to V(\lambda_{V})+S,$
where $S$ denotes the spinless target that retains intact. Definitions of the NPE and UPE helicity amplitudes and formulas for 
the SW SDMEs,  $r_{\lambda_V \tilde{\lambda}_V}^{\alpha}$ 
in terms of the helicity amplitudes can be found in \cite{DC-24}.
Let us denote the experimentally found SDME as $r_{\lambda_V \tilde{\lambda}_V}^{\alpha}$
and the function $f_{\lambda_V \tilde{\lambda}_V}^{\alpha}(F_{\lambda_{V} \lambda_{\gamma}})$ expresses the SDME in terms of the helicity amplitudes,
$F_{\lambda_{V} \lambda_{\gamma}}$. It  may be rewritten through the HARs.
The equations for the helicity-amplitude ratios are
\begin{equation}
f_{\lambda_V \tilde{\lambda}_V}^{\alpha}(F_{\lambda_{V} \lambda_{\gamma}})= r_{\lambda_V \tilde{\lambda}_V}^{\alpha}.
\label{bas-eq}
\end{equation}
The helicity amplitudes, $F_{\lambda_{V} \lambda_{\gamma}}$ for the spinless nucleus  
obey the symmetry relations $F_{-\lambda_{V} -\lambda_{\gamma}}=(-1)^{\lambda_{V}-\lambda_{\gamma}} F_{\lambda_{V} \lambda_{\gamma}}$ 
due to parity conservation in strong and electromagnetic interactions. As follows from these relations 
the number of independent helicity amplitudes is 5, namely $F_{11}=F_{-1-1}$, $F_{10}=-F_{-10}$, $F_{1-1}=F_{-11}$,
$F_{01}=-F_{0-1}$, and $F_{00}$. We choose the amplitude ratios $F_{\lambda_{V} \lambda_{\gamma}}/F_{11}$ since they are finite for any $Q^2$.
The solution of Eqs.~$(\ref{bas-eq})$ is \cite{MSI}
\begin{eqnarray}
 \frac{F_{00}}{F_{11}}=
\frac{\bigl\{r^{04}_{00}+r^{1}_{00}\bigr \} \bigl\{r^{1}_{11}+r^{1}_{1-1}-{\rm{Im}}(r^{2}_{1-1}) -i {\rm{Im}}(r^{3}_{1-1})\bigr \}}
{\epsilon \sqrt{8}\bigl\{-{\rm{Im}}(r^{6}_{10})+i {\rm{Im}}(r^{7}_{10})\bigr \} \bigl\{ r^{1}_{1-1}-{\rm{Im}}(r^{2}_{1-1})\bigr \}},
\;\;\;\frac{F_{1-1}}{F_{11}}=\frac{r^{1}_{11}-i{\rm{Im}}(r^{3}_{1-1})}{r^{1}_{1-1}-{\rm{Im}}(r^{2}_{1-1})},
\label{f00-f11}\\
 \frac{F_{10}}{F_{11}}
=\frac{{\rm{Im}}(r^{6}_{1-1})-r^{5}_{1-1}-i\{r^{8}_{11}-{\rm{Im}}(r^{7}_{1-1}) \}}{\sqrt{2}\bigl\{r^{1}_{1-1}-{\rm{Im}}(r^{2}_{1-1})\bigr\}},
\;\;\;\frac{F_{01}}{F_{11}}=
\frac{2\bigl\{{\rm{Im}}(r^{2}_{10})-i {\rm{Im}}(r^{3}_{10})\bigr\}}{r^{1}_{11}+r^{1}_{1-1}-{\rm{Im}}(r^{2}_{1-1})+i{\rm{Im}}(r^{3}_{1-1})},
\label{f01-f11}
\end{eqnarray}
where $\epsilon$ is the flux ratio of the longitudinally and transversely polarized photons \cite{SW,DC-24}.

It is well known that all 23 SDMEs can be extracted from the angular distribution of the final particles. 
The existence of Eqs.~$(\ref{f00-f11}-\ref{f01-f11})$ means that all four HARs can be obtained from the angular distribution. 
These ratios may be used as free-fit parameters.
Eqs.~$(\ref{f00-f11}-\ref{f01-f11})$
contain those SDMEs  that can be extracted only from the data 
with a polarized beam 
("polarized SDMEs"), hence the application of the amplitude method is possible if the lepton beam is longitudinally polarized. The other condition for the amplitude-method 
applicability is the following: $\epsilon$ should noticeably deviate from 1. 
Indeed, if $\epsilon=1$ the 
contribution of the polarized SDMEs into the angular distribution is zero \cite{SW, DC-24} and  Eqs.~$(\ref{f00-f11}-\ref{f01-f11})$ do not exist.
\section{Virtual-photon Longitudinal-to-transverse Cross-section Ratio}
\label{sect-R}
The ratio of the differential cross sections of the VM production by longitudinally-polarized ($\frac{d \sigma _L}{d t}$) to transversely-polarized
($\frac{d \sigma _T}{d t}$) virtual photons on the spinless target is \cite{MSI}
\begin{equation}
R\equiv \frac{d \sigma _L}{d t}\Big / \frac{d \sigma _T}{d t}
=\frac{r^{04}_{00}+r^{1}_{00}+2(r^{1}_{11}-r^{04}_{1-1})}{\epsilon(2r^{1}_{1-1}-r^{1}_{00})}.
\label{r-less}
\end{equation}
This formula is exact, while all formulas for the VM production on the nucleon expressing
$\widetilde{R}$ in terms of the SDMEs are approximate.
Here, $\widetilde{R}$ instead of $R$ means that $R$ is considered for the nucleon.
The tilde will be used for all quantities for scattering from the nucleon.
The usually applied approximate formula is $\widetilde{R} \approx \widetilde{R}_{04}$, where $\widetilde{R}_{04}
=\widetilde{r}^{04}_{00}/\{\epsilon (1-\widetilde{r}^{04}_{00})\}$ \cite{SW}.
The attempt to use Eq.~$(\ref{r-less})$ for the nucleon is not successful since contributions of the UPE amplitudes into $\widetilde{R}$ are incorrect. The most
important of them is the contribution of $U_{1\frac{1}{2}1\frac{1}{2}}$.
The approximation $\widetilde{R} \approx \widetilde{R}_{D}$ is much better at high energies than $\widetilde{R} \approx \widetilde{R}_{04}$, where \cite{MSI}  
\begin{equation}
\widetilde{R}_D 
=\frac{\widetilde{r}^{04}_{00}+\widetilde{r}^{1}_{00}+2(\widetilde{r}^{1}_{11}-\widetilde{r}^{04}_{1-1})}{\epsilon
\{1-\widetilde{r}^{04}_{00}-\widetilde{r}^{1}_{00}-2(\widetilde{r}^{1}_{11}-\widetilde{r}^{04}_{1-1})\}}.
\label{r-ds-appr}
\end{equation}   
Indeed, corrections to $\widetilde{R}_{D}$ contain the contributions of
only UPE amplitudes excluding $U_{1\frac{1}{2}1\frac{1}{2}}$,
while corrections to $\widetilde{R}_{04}$ contain the contributions both UPE and NPE amplitudes \cite{MSI}. We remind that the ratio of any UPE amplitude to 
any NPE amplitude
tends to zero with increasing of the beam energy. Simple estimates of $\widetilde{R}_D-\widetilde{R}_{04}$ for the HERMES data \cite{DC-24} show
that the difference is of the order of the total uncertainty \cite{MSI}.
Hence the improvement of using $\widetilde{R}_D$ instead of $\widetilde{R}_{04}$ will be important for future precise experiments only.
\section{Non-negativity of angular distribution}
\label{sect-posit}  
The SDMEs should obey positivity-constraint relations in order to make the angular distribution non-negative (see \cite{GT} and references therein). But we should demand more: 
the 
SDMEs
can be represented in terms of the helicity amplitudes. We shall show in this paragraph that for any set of the nonzero helicity amplitudes the angular distribution is 
non-negative. We consider the  $\rho^0$-meson production and its decay into $\pi^+$ and $\pi^-$ but generalizations to other vector mesons are straightforward. The angular 
distribution, 
$\mathcal{W}$ is related to the 
unnormalized spin-density matrix $\Upsilon_{\lambda\tilde{\lambda}}$ of the produced $\rho^0$-meson as
\begin{equation}
\mathcal{N}\mathcal{W}=\sum_{\lambda, \tilde{\lambda}}Y_{1\lambda}(\vartheta, \varphi)Y^*_{1\tilde{\lambda}}(\vartheta, \varphi)
\Upsilon_{\lambda\tilde{\lambda}},
\label{nw-ups}
\end{equation}
where the spherical harmonics $Y_{1\lambda}(\vartheta, \varphi)$ describes the angular part of the decay-pion wavefunction. It depends on the polar 
($\vartheta$) and azimuthal 
($\varphi$) angles of the $\pi^+$ 3-momentum in the rest frame of the $\rho^0$-meson. The normalization factor $\mathcal{N}$ in Eq.~$(\ref{nw-ups})$ is
\begin{equation}
\mathcal{N}=|F_{11}|^2 +|F_{01}|^2+ |F_{-11}|^2+
\epsilon \{|F_{10}|^2+ |F_{00}|^2+|F_{-10}|^2\}.
\label{def-norm}
\end{equation}
The matrix $\Upsilon_{\lambda\tilde{\lambda}}$ is given by the von-Neumann formula
\begin{equation}
\Upsilon_{\lambda_V \tilde{\lambda}_V}=
\sum_{\lambda_{\gamma}, \tilde{\lambda}_{\gamma}}  F_{\lambda_{V}  \lambda_{\gamma}}
   F^*_{\tilde{\lambda}_{V} \tilde{\lambda}_{\gamma}} \varrho_{\lambda_{\gamma} \tilde{ \lambda}_{\gamma}},
\label{upsil}
\end{equation}
where $\varrho_{\lambda_{\gamma} \tilde{ \lambda}_{\gamma}}$ denotes the spin-density matrix of the virtual photon.  The formulas for 
$\varrho_{\lambda_{\gamma} \tilde{ \lambda}_{\gamma}}$ can be found in \cite{SW,DC-24}. Let us consider eigenvalues $\eta_n$ and eigenvectors $X_n$ 
($n=1,\;2,\;3$) of $\varrho_{\lambda_{\gamma} \tilde{ \lambda}_{\gamma}}$.
The formulas for the eigenvalues look like 
\begin{eqnarray}
\eta_1=\frac{1}{2}(1+\epsilon+\sqrt{g}),\;
\eta_2=\frac{1}{2}(1+\epsilon-\sqrt{g}),\; \eta_3=0,
\label{ev-gam}\\
g=4\epsilon^2+P^2_b (1- \epsilon )(1+3 \epsilon)
\label{def-g}
\end{eqnarray}
with $P_b$ being the value of the lepton longitudinal polarization 
($-1 \leq P_b \leq 1$). As follows from
Eq.~$(\ref{def-g})$  $g \geq 0$ for $0 \leq \epsilon \leq 1$, hence $\eta_1 >0$ according to Eqs.~$(\ref{ev-gam})$. As follows from
Eq.~$(\ref{def-g})$
$(1+ \epsilon )^2-g= (1- \epsilon )(1+3 \epsilon) (1-P^2_b) \geq 0.$ Hence
the difference $1+\epsilon-\sqrt{g}=0$ if $P_b^2=1$, or $\epsilon=1$, otherwise it is positive. This means according to Eqs.~$(\ref{ev-gam})$ that $\eta_2 \geq 0$, while
$\eta_3 =0$.

Let us denote the components of the eigenvectors $X_n^{\lambda_{\gamma}}$, then $\varrho_{\lambda_{\gamma}\tilde{\lambda}_{\gamma}}$ is presentable as 
\begin{equation}
\varrho_{\lambda_{\gamma}\tilde{\lambda}_{\gamma}}=
\sum_{n=1}^3X_n^{\lambda_{\gamma}} \eta_n  (X_n^{\tilde{\lambda}_{\gamma}})^* .
\label{rho-eigx}
\end{equation}
Eq.~$(\ref{rho-eigx})$ is valid since 
$\varrho_{\lambda_{\gamma}\tilde{\lambda}_{\gamma}}$ is Hermitian, hence the vectors $X_n$ are mutually orthogonal
and normalized to unity. Substituting Eq.~$(\ref{rho-eigx})$ into Eq.~$(\ref{upsil})$ and Eq.~$(\ref{upsil})$ into Eq.~$(\ref{nw-ups})$ one gets 
\begin{equation}
\mathcal{N} \mathcal{W}=\sum_{n=1}^3|\mathcal{G}_{n}|^2\eta_n\equiv \sum_{n=1}^3\Bigl | \sum_{\lambda_V,\lambda_{\gamma}}Y_{1\lambda_V}
F_{\lambda_V\lambda_{\gamma}}X_n^{\lambda_{\gamma}}\Bigr |^2 \eta_n.
\label{diag-w}
\end{equation}
Since $\eta_1 > 0$, $\eta_2\geq 0$,  $\eta_3=0$, $|\mathcal{G}_{n}|^2 \geq 0$ 
and $\mathcal{N}>0$ in Eq.~$(\ref{diag-w})$ if at least one amplitude in Eq.~$(\ref{def-norm})$
is nonzero, then $\mathcal{W}\geq 0$. This means that the angular distribution is non-negative
even if amplitudes do not describe the VM production.
This property is important for the maximum-likelihood-method application
 in the data description by the amplitude method.
\section{Constraint relations}
\label{sect-constr}
The total number of the SW SDMEs 
is 23. The number of the complex HARs is 4
(8 real functions).
Hence $23-8=15$ QM-constraint relations exist. The abbreviation QM reminds that in quantum mechanics (QM) the SDMEs are expressible via 
 the amplitudes. Since all the UPE amplitudes for spinless targets are zero there are 3 linear QM-constraints  
$u_1=1-r^{04}_{00}+2r^{04}_{1-1}-2r^{1}_{11}-2r^{1}_{1-1}=0,\;\;u_2=r^{5}_{11}+r^{5}_{1-1}=0,\;\;u_3=r^{8}_{11}+r^{8}_{1-1}=0.$
For the nucleon targets, $u_1,\;u_2,\;u_3$ are nonzero and show the contribution of the UPE amplitudes to the SDMEs \cite{SW,DC-24}.
All other 12 independent QM-constraint relations are nonlinear \cite{MSI}. 

The angular distribution becomes distorted by any real detector. The original distribution can be restored using Monte Carlo (MC) 
simulation of the detector property (unfolding). 
If the MC codes describe the detector property not well enough the extracted SDMEs have appreciable systematic uncertainties.
A comparison of the SDMEs obtained in the SDME method with the SDMEs calculated with the HARs got in the 
amplitude method estimates the quality of the MC codes. A check of QM-constraint-relation validity does the same. 
\vspace{-0.25cm}
\section{Conclusion}
\label{sect-concl}
For the VM production with polarized beams on spinless targets, there are the relations expressing all the HARs through the  
SDMEs. This permits to apply the amplitude method considering the HARs as free-fit parameters.
The advantages of the amplitude method in respect of the SDME one  are as follows. First, the number of free-fit parameters in the former is 8 and 23 
 in the latter. Second, the angular distribution is non-negative for any set of non-zero amplitudes, while the SDMEs must obey the positivity-constrain 
relations and the QM-constrain relations. The latter are valid for the amplitude method by default. A comparison of the SDMEs obtained 
in the SDME method and calculated in the amplitude method permits to estimate the systematic uncertainty of the applied unfolding procedure. 

The exact formula for the virtual-photon longitudinal-to-transverse cross-section ratio for spinless targets and the approximate formula for nucleons obtained in \cite{MSI} are 
discussed. 
\begin{acknowledgments}
I am much obliged to all participants of 
XIX Workshop on High Energy Spin Physics in Dubna (DSPIN-23) for many fruitful discussions and especially to Prof. O. Teryaev who has drawn my attention
to some publications devoted to the subject under consideration.
\end{acknowledgments}


\begin{thebibliography}{100}
\bibitem{INS} I.~P.~Ivanov, N.~N.~Nikolaev, A.~A.~Savin, Phys. Part. Nucl. {\bf 37}, 1 (2006).
\bibitem{R83} M.~G.~Ryskin, Z. Phys. C {\bf 57}, 89 (1993).
\bibitem{R99} M.~G.~Ryskin, Phys. Atom. Nucl. {\bf 62}, 315 (1999).
\bibitem{SW} K.~Schilling, G.~Wolf, Nucl. Phys. B {\bf 61}, 381 (1973).
\bibitem{AMA} F.~D.~Aaron et al. (H1 Collab.), JHEP {\bf 1005}, 32 (2010).
\bibitem{DC-84} A.~Airapetian et al. (HERMES Collab.), Eur. Phys. J. C {\bf 71}, 1609 (2011).
\bibitem{ABB} J.~J.~Aubert et al. (EMC Collab.), Phys. Lett. B {\bf 123}, 275 (1983).
\bibitem{HMPW} O.~Hen, G.~A.~Miller, E.~Piasetzky, L.~B.~Weinstein, Rev. Mod. Phys. {\bf 89}, 045002 (2017).
\bibitem{DC-24} A.~Airapetian et al. (HERMES Collab.), Eur. Phys. J. C {\bf 62}, 659 (2009).
\bibitem{MSI} S.~I.~Manaenkov, Phys. Part. Nucl. Lett. {\bf 21}, 31 (2024).
\bibitem{GT} M.~Gavrilova, O.~Teryaev, Phys. Rev. D {\bf 99}, 076013 (2019).
\end{thebibliography}
\end{document}